\documentclass[cameraready]{Interspeech}

\title{On the Distillation Loss Functions of Speech VAE for Unified Reconstruction, Understanding, and Generation}

\author[affiliation={1}, orcid=0009-0002-2867-3699]{Changhao}{Cheng}
\author[affiliation={2},orcid=0009-0006-9498-1592]{Wei}{Wang}
\author[affiliation={1},orcid=0000-0003-4500-3515]{Wangyou}{Zhang}
\author[affiliation={3},orcid=0009-0000-3204-3759]{Dongya}{Jia}
\author[affiliation={3},orcid=0000-0002-3393-1130]{Jian}{Wu}
\author[affiliation={3}]{Zhuo}{Chen}
\author[affiliation={2},orcid=0000-0002-0314-3790]{Yanmin}{Qian}

\address{
    $^1$ School of Artificial Intelligence, Shanghai Jiao Tong University, Shanghai, China \\
    $^2$ Auditory Cognition and Computational Acoustics Lab\\ MoE Key Lab of Artificial Intelligence, AI Institute\\
School of Computer Science, Shanghai Jiao Tong University, Shanghai, China\\
$^3$ ByteDance Seed, China
}

\email{taycch1213@sjtu.edu.cn}


\usepackage{comment}
\usepackage{booktabs} 
\usepackage{multirow} 
\usepackage{xcolor}     
\usepackage{colortbl}
\usepackage{graphicx}    
\usepackage{amsmath}     
\usepackage{makecell} 
\usepackage{booktabs}
\usepackage{subcaption}
\usepackage{hyperref} 
\usepackage{caption} 
\usepackage{arydshln}

\begin{document}

\maketitle

\begin{abstract}
Continuous speech representations based on Variational Autoencoders (VAEs) have emerged as a promising alternative to traditional spectrogram or discrete token based features for speech generation and reconstruction. Recent research has tried to enrich the structural information in VAE latent representations by aligning with self-supervised learning (SSL) features, aiming for better generation performance. However, it remains unclear whether the widely-used alignment approach based on time-axis distillation is optimal when considering more tasks. To address this problem, this paper systematically explores different alignment approaches and analyzes their impact on the performances over three axes: reconstruction, understanding, and generation. We investigate various design choices in the distillation loss. Extensive experiments show that the joint-marginal alignment approach with adaptive weighting can achieve the best overall performance while allowing for a controllable balance.

\end{abstract}

\vspace{-6pt}
\section{Introduction}
\vspace{-3pt}
Discrete speech representations, including both semantic tokens (e.g.,  HuBERT \cite{hsu2021hubert}) and acoustic tokens (e.g., neural audio codecs \cite{zeghidour2021soundstream, defossez2022high, yang2023hifi}), have proven effective for boosting the performance of speech large language models (Speech LLMs) \cite{casanova2025low, ye2025codec}. However, quantizing continuous audio signals into discrete tokens incurs inevitable information loss, which has motivated growing interest in continuous representations learned via deep learning like Variational Autoencoders (VAEs) \cite{xia2024kall, jia2025ditar, wu2025clear, peng2025vibevoice, zhou2025voxcpm}.
Consequently, continuous speech representations such as VAE latent features have been shown to provide rich information for speech generative modeling while keeping compactness~\cite{jia2025ditar}.

In the literature, VAE is commonly observed to suffer from the reconstruction-loss trade-off \cite{gupta2024photorealistic}.
To address this problem, the VA-VAE approach \cite{yao2025reconstruction} has been proposed to align the VAE latent space with intermediate features from vision foundation models.
A similar method has also been explored for speech: Semantic-VAE~\cite{niu2025semantic} uses a time-axis (T-axis) point-wise cosine similarity loss to distill the features of speech foundation models into speech VAE latent space, outperforming both mel spectrogram and Vanilla VAE in the zero-shot text-to-speech (TTS) task~\cite{chen2025f5}.

\begin{figure*}
    \centering
    \includegraphics[width=0.95\linewidth]{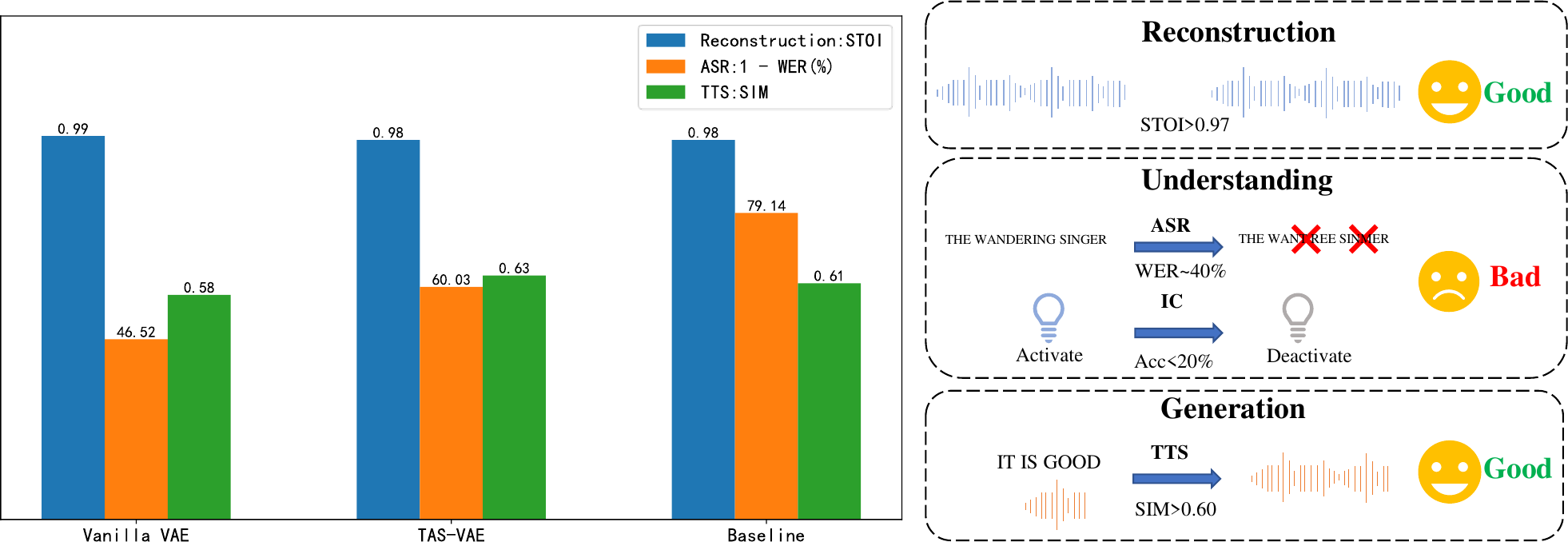}
    \caption{T-axis Aligned Semantic VAE (TAS-VAE) distills semantic knowledge from speech foundation models via Eq.\ref{T-axis loss} alignment loss, achieving TTS performance comparable to mel spectrograms with minor reconstruction degradation. Its latent representations still underperform on downstream speech understanding. Baseline: Mel+Vocos~\cite{siuzdakvocos} (reconstruction), Fbank (understanding), Mel+F5-TTS~\cite{chen2025f5} (generation).}
    \label{fig:dilemma of speech vae}
    \vspace{-10pt}
\end{figure*}

However, existing studies mainly optimize VAEs for reconstruction or generation tasks, neglecting other important aspects of speech research, i.e., understanding (e.g., speech recognition, speaker verification, emotion classification, etc.).
In fact, our experiments show that although the alignment scheme of T-axis Aligned VAE (TAS-VAE) improves both generation and understanding over Vanilla VAE, its understanding capability remains poor.
For example, its Automatic Speech Recognition (ASR) performance in terms of Word Error Rate (WER) is much higher than that of Fbank features (cf. Fig. \ref{fig:dilemma of speech vae}).

As recent trends move toward unifying speech understanding and generation models \cite{huang2025step, wang2025dualspeechlm}, it is crucial to develop acoustic representations that serve both purposes effectively. We argue that the dilemma illustrated in Fig.~\ref{fig:dilemma of speech vae} arises from the fact that conventional distillation loss functions overlook the structural differences inherent in representations. Thus, by introducing a novel distillation paradigm, we have identified a speech VAE representation that unifies and better balances reconstruction, generation, and understanding. Overall, we not only present a novel VAE approach that outperforms traditional continuous representations across various downstream tasks,  but also provide detailed insights into tuning the trade-offs between these downstream capabilities. We also release models and code to facilitate further research \footnote{https://github.com/changhao-cheng/JMAS-VAE}.

\vspace{-6pt}
\section{Proposed Methods}
\begin{figure}[h]
    \centering
    \includegraphics[width=0.95\linewidth]{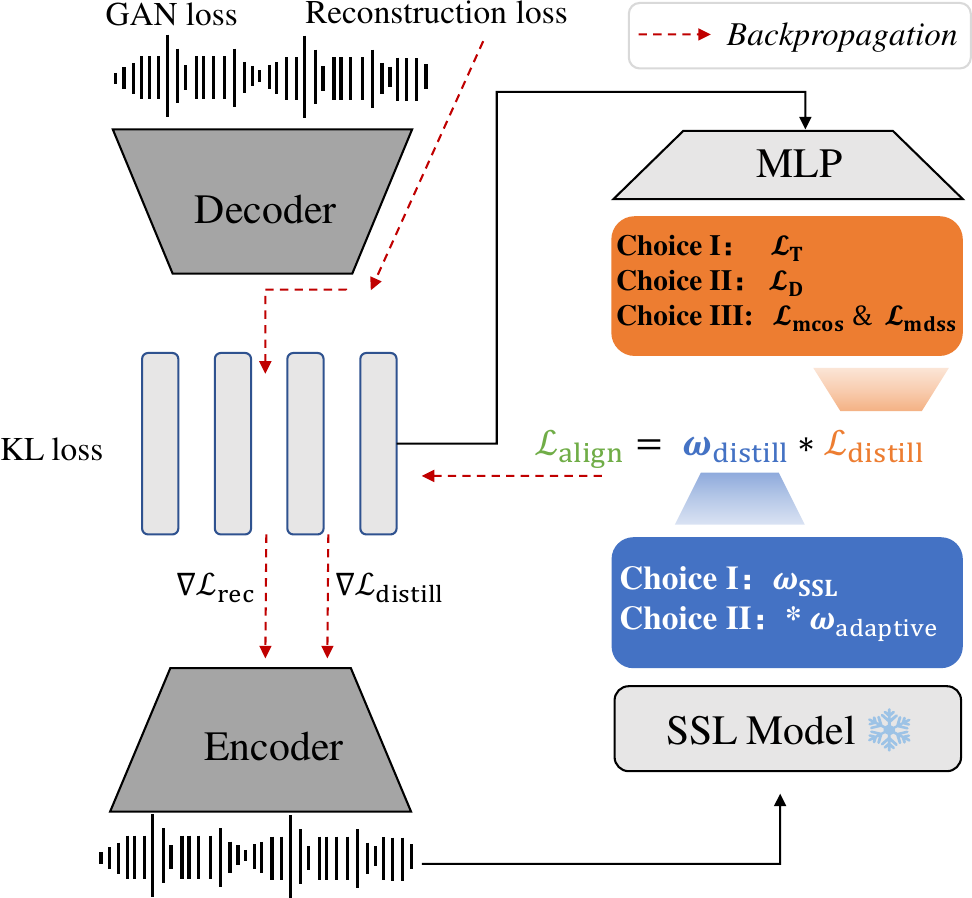}
    \caption{\textbf{The design space of distillation loss functions for speech VAEs.} For mathematical formulations, see Eqs. \ref{T-axis loss} to \ref{adaptive loss}.
}
    \label{fig:vae structure}
\end{figure}

\vspace{-3pt}
The widely adopted loss combination for VAE training is illustrated in Fig.~\ref{fig:vae structure}, which consists of a reconstruction loss for autoencoding, a Kullback-Leibler (KL) divergence loss for posterior regularization, and Generative Adversarial Network (GAN) based losses for distribution matching~\cite{StableAudioOpen}.
Following Semantic-VAE~\cite{niu2025semantic}, we also include a feature alignment loss $\mathcal{L}_{\mathrm{align}}$ on the VAE latent, as shown in the right part of Fig.~\ref{fig:vae structure}.

Compared to conventional VAEs, the core modification lies in aligning with the foundation models, such as self-supervised learning (SSL) models based on a distillation loss term:
\begingroup  
\setlength{\abovedisplayskip}{2pt} 
\setlength{\belowdisplayskip}{2pt}   
\setlength{\abovedisplayshortskip}{2pt} 
\setlength{\belowdisplayshortskip}{2pt}  
\begin{equation}
    \mathcal{L}_{\mathrm{align}} = \omega_{\mathrm{distill}}\mathcal{L}_{\mathrm{distill}}\,,
    \label{L_ssl}
\end{equation}
\endgroup  
where $\omega_{\mathrm{distill}}$ is the loss weight factor, and $\mathcal{L}_{\mathrm{distill}}$ is a loss function used to distill structural semantic information from pretrained SSL models into VAE latent representations.
Our goal is to explore different design choices regarding this loss term and analyze the corresponding impact on downstream performances.
\vspace{-6pt}
\subsection{Alignment Loss Function Design Space}
\vspace{-6pt}
\subsubsection{T-axis Aligned Semantic VAE}
\vspace{-2pt}
The mathematical form of $\mathcal{L}_{\mathrm{distill}}$ exerts a notable influence on the downstream performances of speech VAEs.
A commonly-used scheme is to align the features with T-axis cosine distance loss, which is shown to outperform mean absolute error (MAE) and mean squared error (MSE) losses in the speech generation task~\cite{niu2025semantic}.
Let $f$ denote the speech feature extracted by the SSL model, and $z'$ the dimensionally aligned VAE latent obtained via multilayer perceptron (MLP) projection. The T-axis distillation loss can be then expressed as:
\begingroup  
\setlength{\abovedisplayskip}{2pt} 
\setlength{\belowdisplayskip}{2pt}   
\setlength{\abovedisplayshortskip}{2pt} 
\setlength{\belowdisplayshortskip}{2pt}  
\begin{equation}
    \mathcal{L}_{\mathrm{T}} = -\frac{1}{BT}\sum_{b=1}^{B}\sum_{t = 1}^{T}\log\sigma(\cos({z'_{b,t}, f_{b,t}}))\,,
    \label{T-axis loss}
\end{equation}
\endgroup
where $T$ denotes the total number of time frames, $t$ the time step, $B$ the batch size, $b$ the sample index and $\sigma(\cdot)$ the sigmoid activation.

\vspace{-6pt}
\subsubsection{D-axis Aligned Semantic VAE}
\vspace{-2pt}
Orthogonal to point-wise temporal alignment, we further introduce the dimension-wise distillation scheme, which captures the temporal structure and variations in local feature dimensions: 
\begingroup  
\setlength{\abovedisplayskip}{2pt} 
\setlength{\belowdisplayskip}{2pt}   
\setlength{\abovedisplayshortskip}{2pt} 
\setlength{\belowdisplayshortskip}{2pt}  
\begin{equation}
    \mathcal{L}_{\mathrm{D}} = -\frac{1}{BD}\sum_{b = 1}^{B}\sum_{d = 1}^{D}\log\sigma(\cos({z'^{[d]}_{b}, f^{[d]}_b})),
    \label{D-axis loss}
\end{equation}
\endgroup
where $D$ is the total feature dimensionality of $z'$ and $f$, and $d$ represents the dimensional index. Similar distillation schemes have been employed in discrete spaces and demonstrated promising performance \cite{zhang2023speechtokenizer}.

\begin{table*}[htbp]
  \centering
  \fontsize{8.2}{9.0}\selectfont
  \setlength{\tabcolsep}{2pt}
  \renewcommand{\arraystretch}{1.2}
  \caption{Performance comparison of different speech representation methods. For methods, $^\star$ denotes the use of adaptive weighting in Section~\ref{ssec:loss_weight}. The metrics for the eight understanding tasks and the WER metric in the generation task are all percentages.}
  \label{tab:method_performance}
  \begin{tabular}{l|cc>{\columncolor{gray!20}}c|cccccccc>{\columncolor{gray!20}}c|cc>{\columncolor{gray!20}}c|c}
    \hline
    \multirow{3}{*}{\textbf{Method}} 
    & \multicolumn{3}{c|}{\multirow{2}{*}{\textbf{Recon.}}} 
    & \multicolumn{9}{c|}{\textbf{Under.}} 
    & \multicolumn{3}{c|}{\multirow{2}{*}{\textbf{Gene.}}} 
    & \multirow{3}{*}{\makecell{\textbf{Overall}\\ \textbf{Score $\uparrow$}}} \\
    & \multicolumn{3}{c|}{} 
    & \textbf{ER} & \textbf{PR} & \textbf{ASR} & \textbf{KS} & \textbf{SID} & \textbf{ASV} & \textbf{SD} & \textbf{IC} & \cellcolor{white!20}
    & \multicolumn{3}{c|}{} & \\
    \cline{2-16}
    & \textbf{PESQ$\uparrow$} & \textbf{STOI$\uparrow$} & \textbf{$x_r\uparrow$} 
    & \textbf{Acc$\uparrow$} & \textbf{PER$\downarrow$} & \textbf{WER$\downarrow$} 
    & \textbf{Acc$\uparrow$} & \textbf{Acc$\uparrow$} & \textbf{EER$\downarrow$} & \textbf{DER$\downarrow$} & \textbf{Acc$\uparrow$} & \textbf{$x_u\uparrow$} 
    & \textbf{WER$\downarrow$} & \textbf{SIM$\uparrow$} & \textbf{$x_g\uparrow$} &\\
    \hline
    Vanilla VAE         & \textbf{4.12} & \textbf{0.985} & \textbf{0.905} &36.87 & 89.40 & 53.48 & 29.80 & 7.74 & 14.64 & 17.11 & 5.98 & 0.382 & 2.72 & 0.58 & 0.776 & 0.645\\
    Semantic-VAE~\cite{niu2025semantic}        & 3.97 & 0.981& 0.888& 45.99 & 85.71 & 41.75 & 43.49 & 15.97& 12.41 & 15.21 & 9.25& 0.449& 2.01 & \textbf{0.67} & \textbf{0.825} & 0.690\\
    EnCodec~\cite{defossez2022high}            & 2.77 & 0.938 & 0.746 & 50.41 & 80.75 & 28.90 & 48.46 & 18.08 & 12.05 & 14.10 & 10.04 & 0.489 &4.71 & 0.56 & 0.756 &0.651\\
    Baseline (Mel/Fbank) & 3.60 & 0.978& 0.849 & 35.39 & 82.01 & 20.86 & 8.63 & 8.5E-4 & 9.56 & 10.05 & 9.10 & 0.413 &2.23 & 0.61 & 0.794 &0.653\\
    \hdashline
    TAS-VAE                   & 3.97 & 0.977 & 0.886 & 44.61 & 70.12 & 39.97 & 64.56 & 17.65 & 11.67 & 13.64 & 16.93 & 0.510 &2.55 & 0.63& 0.802 & 0.713\\
    TAS-VAE$^\star$           & 2.92 & 0.947& 0.766 & 56.77 & 19.70 & 15.40 & \textbf{96.62} & 23.11 & \textbf{8.80} & \textbf{9.93} & 71.74& 0.743& \textbf{1.93} & 0.31 & 0.645 & 0.716\\
    DAS-VAE                   & 3.98 & 0.979 & 0.888 & 53.73 & 52.84 & 27.83 & 80.46 & 19.92 & 10.61 & 12.54 & 28.55 & 0.599 &2.46 & 0.59& 0.783& 0.746\\
    DAS-VAE$^\star$           & 2.73 & 0.940& 0.743 & \textbf{60.18} & \textbf{19.66} & \textbf{14.86} & 96.36 & 22.42 & 8.83 & 10.68 & \textbf{76.22} & \textbf{0.751} &2.32 & 0.32 & 0.648 & 0.713\\
    JMAS-VAE      & 3.97 & 0.978 & 0.886 & 46.18 & 70.95 & 37.49 & 61.57 & 17.61 & 11.95 & 12.81 & 18.11& 0.513 & 2.59 & 0.63 & 0.802 &0.714\\
    JMAS-VAE$^\star$ & 3.84 & 0.973& 0.871 & 57.24 & 36.72 & 21.04 & 92.76 & \textbf{24.58} & 9.53 & 10.65 & 48.48 & 0.681 &2.04 & 0.57 & 0.775&\textbf{0.772}\\
    \hline
  \end{tabular}
  \vspace{-16pt}
\end{table*}

\vspace{-6pt}
\subsubsection{Joint-marginal Aligned Semantic VAE}
\label{ssec:jmas_vae}
\vspace{-2pt}
In addition to fine-grained frame-wise and dimension-wise alignment, we further incorporate joint-marginal distribution alignment into our design space.
In contrast to existing point-wise losses that only focus on overall feature matching, this scheme further enforces the intra-distribution consistency between the latent space and the foundation model feature space via an extra alignment loss term.

Targeting the long-range dependence of speech sequences, we adapt the joint-marginal alignment loss from the computer vision domain \cite{yao2025reconstruction} to speech representation processing. We construct the loss at two levels: frame-level feature distance and sequence-level distribution similarity. The marginal cosine similarity loss aligns SSL and VAE features along the T-axis:
\begingroup
\setlength{\abovedisplayskip}{2pt}
\setlength{\belowdisplayskip}{2pt}
\setlength{\abovedisplayshortskip}{2pt}
\setlength{\belowdisplayshortskip}{2pt}
\begin{equation}
    \mathcal{L}_{\mathrm{mcos}} = \frac{1}{BT}\sum_{b=1}^B\sum_{t = 1}^{T}\mathrm{ReLU}(1 - m_1 - \cos({z'_{b,t}, f_{b,t}}))\,,
    \label{mcos}
\end{equation}
\endgroup
and the marginal distance sequence similarity loss further aligns their relative structures by comparing all pairs of frames:
\begingroup  
\setlength{\abovedisplayskip}{4pt}
\setlength{\belowdisplayskip}{4pt} 
\begin{equation}
\resizebox{0.85\linewidth}{!}{%
  $\displaystyle
  \mathcal{L}_{\mathrm{mdss}} =
  \frac{1}{(BT)^2}
  \sum_{i, j}^{BT}
  \mathrm{ReLU}\big(
  |\cos(z'_i,z'_j)-\cos(f_i,f_j)|-m_2
  \big)$
}
\label{mdss}
\end{equation}
\endgroup
In Eqs \ref{mcos} and \ref{mdss}, $m_1$ and $m_2$ denote the respective margins, which serve as an important mechanism for balancing reconstruction, understanding, and generation capabilities.
\vspace{-10pt}
\subsection{Loss Weight Design Space}
\label{ssec:loss_weight}
\vspace{-4pt}
Since VAE is typically trained in a multi-task manner, the relative weighting of different loss terms, especially the alignment losses, can lead to significant performance variations.
Apart from static loss weights~\cite{niu2025semantic} that need manual tuning, we further investigate an adaptive weighting strategy as follows:
\begingroup  
\setlength{\abovedisplayskip}{2pt} 
\setlength{\belowdisplayskip}{2pt}   
\setlength{\abovedisplayshortskip}{2pt} 
\setlength{\belowdisplayshortskip}{2pt}  
\begin{equation}
    \omega_{\mathrm{adaptive}} = \frac{\lVert \mathcal{\nabla L_{\mathrm{rec}}} \rVert}{\lVert \mathcal{\nabla L_{\mathrm{distill}}}\rVert}\,,
    \label{adaptive loss}
\end{equation}
\endgroup
which dynamically adjusts the loss weight in Eq.~\ref{L_ssl} based on the gradient norm ratio between the reconstruction loss $\mathcal{L}_{\mathrm{rec}}$ and the alignment loss $\mathcal{L}_{\mathrm{distill}}$ (Both loss gradient values are calculated on the parameters of the projection layer of the Encoder.).
Specifically, we have $\omega_{\mathrm{distill}}\in\{\omega_{\mathrm{SSL}}, \omega_{\mathrm{SSL}}*\omega_{\mathrm{adaptive}}\}$, where $\omega_{\mathrm{SSL}}$ is the static weight set manually.
To precisely allocate the contributions of $\mathcal{L}_{\mathrm{mcos}}$ and $\mathcal{L}_{{\mathrm{mdss}}}$ in the joint marginal loss, we compute the adaptive weights separately for each term:
\begingroup  
\setlength{\abovedisplayskip}{2pt} 
\setlength{\belowdisplayskip}{2pt}   
\setlength{\abovedisplayshortskip}{2pt} 
\setlength{\belowdisplayshortskip}{2pt}  
\begin{equation}
    \omega_{\mathrm{mcos}} = \frac{\lVert \mathcal{\nabla L_{\mathrm{rec}}} \rVert}{\lVert \mathcal{\nabla L_{\mathrm{mcos}}}\rVert}, \omega_{\mathrm{mdss}} = \frac{\lVert \mathcal{\nabla L_{\mathrm{rec}}} \rVert}{\lVert \mathcal{\nabla L_{\mathrm{mdss}}}\rVert}.
\end{equation}
\endgroup

\vspace{-6pt}
\section{Experimental Setup}
\vspace{-3pt}
We use \texttt{stable-audio-tools}\footnote{\url{https://github.com/Stability-AI/stable-audio-tools}} as the speech VAE backbone, with a DAC-based encoder~\cite{kumar2023high} and a BigVGAN decoder~\cite{lee2023bigvgan}. The input speech signal is downsampled by factors of \{4,4,5,5\} to a 64-dim 40 Hz latent representation $z$, which is linearly projected to 1024-dim ($z'$) and aligned with the 23rd-layer features of WavLM Large \cite{chen2022wavlm}.

All VAE models are trained on the full Libriheavy \cite{kang2024libriheavy} (16kHz). For the Vanilla VAE, we train it with a batch size of 20 for 550k steps. For TAS‑VAE and DAS‑VAE (DAS‑VAE) with adaptive weighting, we adopt 8 GPUs with a per-GPU batch size of 2 and train for 1100k steps. All other VAE experiments use the same GPU and batch setup and are trained for 600k steps. Other hyperparameters are set as follows: Adam optimizer with $\mathrm{lr}=10^{-4}$ and $\gamma=0.999996$; the loss weights are $\omega_{\mathrm{rec}}=1.0$, $\omega_{\mathrm{KL}}=0.001$, and $\omega_{\mathrm{SSL}}=2.5$.

We evaluate the performance across three categories of downstream tasks:
(1) Speech reconstruction ability is evaluated on LibriSpeech-test-clean \cite{panayotov2015librispeech} with PESQ (Perceptual Evaluation o Speech Quality) \cite{rix2001perceptual} and STOI (Short-timeObjective Intelligibility).
(2) Speech understanding ability is tested on 8 SUPERB \cite{SUPERB-Interspeech2021} tasks: Emotion Recognition (ER), Phoneme Recognition (PR), ASR, Keyword Spotting (KS), Speaker Identification (SID), Automatic Speaker Verification (ASV), Speaker Diarization (SD) and Intent Classification (IC). To obtain results closer to convergence, we increased the training steps from 200k to 500k for all methods (including the baseline) on both the ASR and SID tasks. Consequently, the ASR performance of Fbank in our experimental results is slightly higher than that reported in SUPERB \cite{SUPERB-Interspeech2021}.
(3) We assess the speech generation ability on the TTS task. A small F5-TTS \cite{chen2025f5} model is trained on LibriTTS \cite{zen2019libritts} with a per-GPU batch size of 6400 over 8 GPUs and a learning rate of $10^{-4}$ (For Encodec, gradient explosion would occur under the same learning rate, so we train F5-TTS under this representation with a learning rate of $10^{-5}$.). It is evaluated on LibriSpeech-PC test-clean \cite{meister2023librispeech} using WER and speaker similarity (SIM) metrics.

\vspace{-6pt}
\section{Results}
\vspace{-3pt}
\subsection{Overall Performance Comparison}
\label{ssec:exp_overall}
\vspace{-2pt}

Tab. \ref{tab:method_performance} presents the evaluation results of speech VAEs with different distillation schemes on reconstruction, understanding, and generation. For the joint-marginal-aligned semantic-VAE (JMAS-VAE) in Section~\ref{ssec:jmas_vae}, we set $m_1 = 0.5$ and $m_2 = 0.25$. Besides Vanilla VAE, we compare with Semantic-VAE \cite{niu2025semantic} (whose distillation scheme corresponds to TAS-VAE in this work) and EnCodec (24kHz) \cite{defossez2022high}, and use conventional continuous representation as the baseline. For the \textbf{baseline}, reconstruction is evaluated based on the mel-based vocoder (Vocos \cite{siuzdakvocos}), understanding by Fbank features, and generation by mel-based F5-TTS.

To compute the overall score, we first take the arithmetic mean for each task: reconstruction-$x_r$, understanding-$x_u$ and generation-$x_g$ (Error rates are replaced with ($1-$error rate), and PESQ is rescaled to a similar score scale by PESQ$/5$):
{\setlength{\abovedisplayskip}{4pt} 
\setlength{\belowdisplayskip}{4pt}  
\begin{equation}
\begin{gathered}
    x_r = \frac{\text{PESQ}/5+\text{STOI}}{2},\quad x_g = \frac{1 - \text{WER} + \text{SIM}}{2}\,,\\
    x_u = \frac{\sum_{\text{ER, KS, SID, IC}}\text{Acc}+\sum_{\text{PR, ASR, ASV, SD}}(1-\text{ER})}{8}\,.
\end{gathered}
\end{equation}
}To avoid biased evaluation by extreme scores in some tasks, the overall score in Tab.~\ref{tab:method_performance} is obtained via the geometric mean\footnote{We also explored arithmetic and harmonic means for overall score calculation (cf.\url{https://github.com/changhao-cheng/JMAS-VAE/blob/main/VAE_scores.pdf}), showing a similar conclusion.}~\cite{fleming1986not} $\sqrt[\uproot{2}3]{x_r \cdot x_u \cdot x_g}$.
Comparing the overall scores in Tab.~\ref{tab:method_performance}, we have the following findings:
\begin{enumerate}
    \item Although the Vanilla VAE and Semantic-VAE excel in reconstruction and generation, their performances across eight speech understanding tasks are very poor, some of which even lagging behind the conventional baseline and Encodec.
    \item Compared to TAS-VAE, DAS-VAE demonstrates much better understanding performance with small performance drop in generation, resulting in substantial overall improvement.
    \item The JMAS-VAE model with adaptive weighting significantly outperforms other approaches in terms of the overall score. This highlights the efficacy of the joint-marginal alignment in balancing reconstruction, understanding, and generation within compact continuous representations.
    \item The adaptive weighting strategy (denoted with $^\star$) can significantly improve speech understanding performance for all three types of semantic-aligned VAEs.
    Among them, the JMAS-VAE achieves the best balance of reconstruction, understanding, and generation, while TAS-VAE and DAS-VAE significantly sacrifices their reconstruction and generation abilities when applying adaptive weighting.
\end{enumerate}

The above observations also indicate that directly aligning VAE latents with pretrained SSL features without dedicated regularization can potentially lead to biased representations.
\vspace{-6pt}
\subsection{Analysis of Adaptive Weighting}
\label{ssec:exp_adaptive_weighting}
\vspace{-2pt}

\begin{figure}[htbp]
     \centering 
    \begin{subfigure}{0.22\textwidth}
        \centering 
        \includegraphics[width=\linewidth]{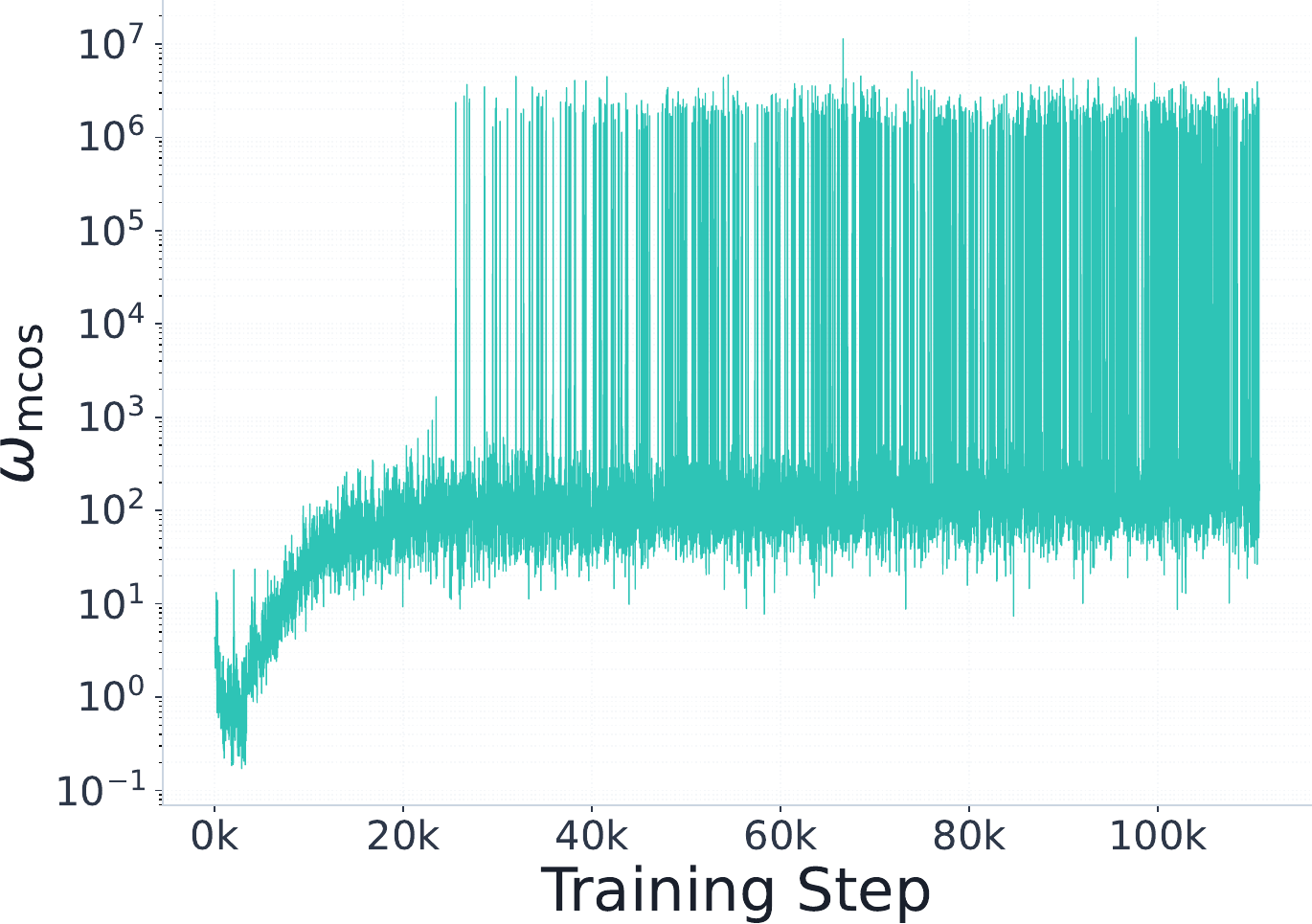}
        \subcaption{Adaptive weighting of $\mathcal{L}_{\text{mcos}}$} 
        \label{subfig:1}      
    \end{subfigure}
    \hfill 
    \begin{subfigure}{0.22\textwidth}
        \centering
        \includegraphics[width=\linewidth]{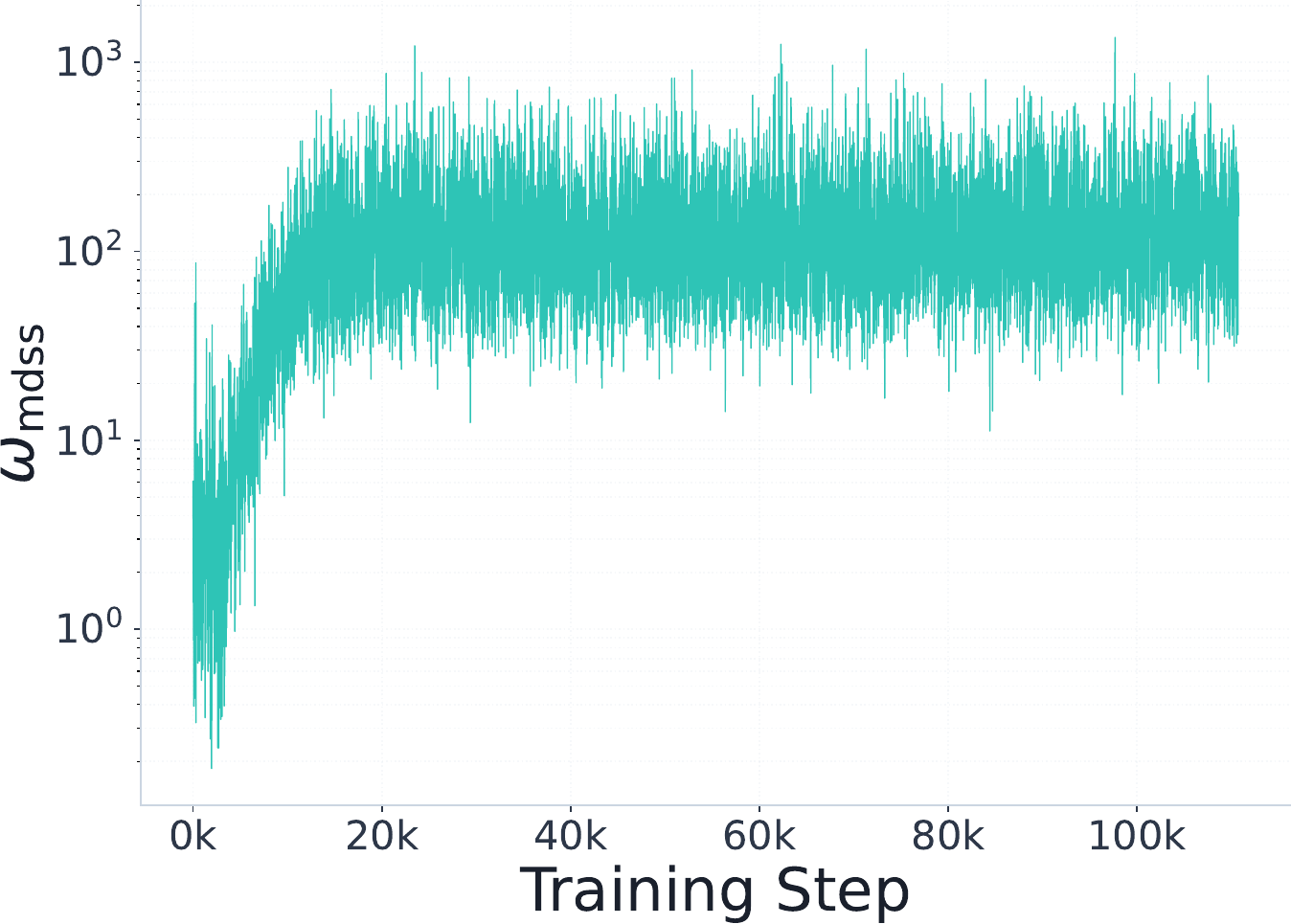}
        \subcaption{Adaptive weighting of $\mathcal{L}_{\text{mdss}}$}
        \label{subfig:2}
    \end{subfigure}
    \caption{The adaptive weighting of the two distillation losses for JMAS-VAE as a function of training steps. The vertical axis adopts a logarithmic coordinate system.} 
    \label{fig:adap_curve}
    \vspace{-6pt}
\end{figure}

As shown in last subsection, the adaptive weighting proposed in Section~\ref{ssec:loss_weight} is critical for better distilling the semantic information structure inherent in pretrained SSL features, thus greatly improving the understanding ability.
In this subsection, we further delve into the adaptive weighting procedure by visualizing the evolving loss weights during training.
As shown in Fig. \ref{fig:adap_curve}, the adaptive weights $\omega_{\text{mcos}}$ and $\omega_{\text{mdss}}$ for the joint-marginal alignment loss quickly become orders of magnitude larger than the commonly-adopted static weights (e.g., $<10$), enabling finer alignment between learned representations and SSL features.

\vspace{-6pt}
\subsection{Ablation Study of JMAS-VAE}
\label{ssec:exp_jmas_vae}
\vspace{-2pt}
The results in Tab. \ref{tab:method_performance} show that simply aligning features toward those of the speech SSL model only improves performance on speech understanding. For speech reconstruction and generation, however, such alignment requires an appropriate threshold—either by disabling adaptive weighting or by manually introducing margins as in Eq.~\ref{mcos}--\ref{mdss}.
Here, we aim to investigate the role of different margins introduced in Section~\ref{ssec:jmas_vae} by training and evaluating different variants of JMAS-VAE.

\begin{figure}[htbp] 
    \centering 
    \begin{subfigure}{0.22\textwidth}
        \centering 
        \includegraphics[width=\linewidth]{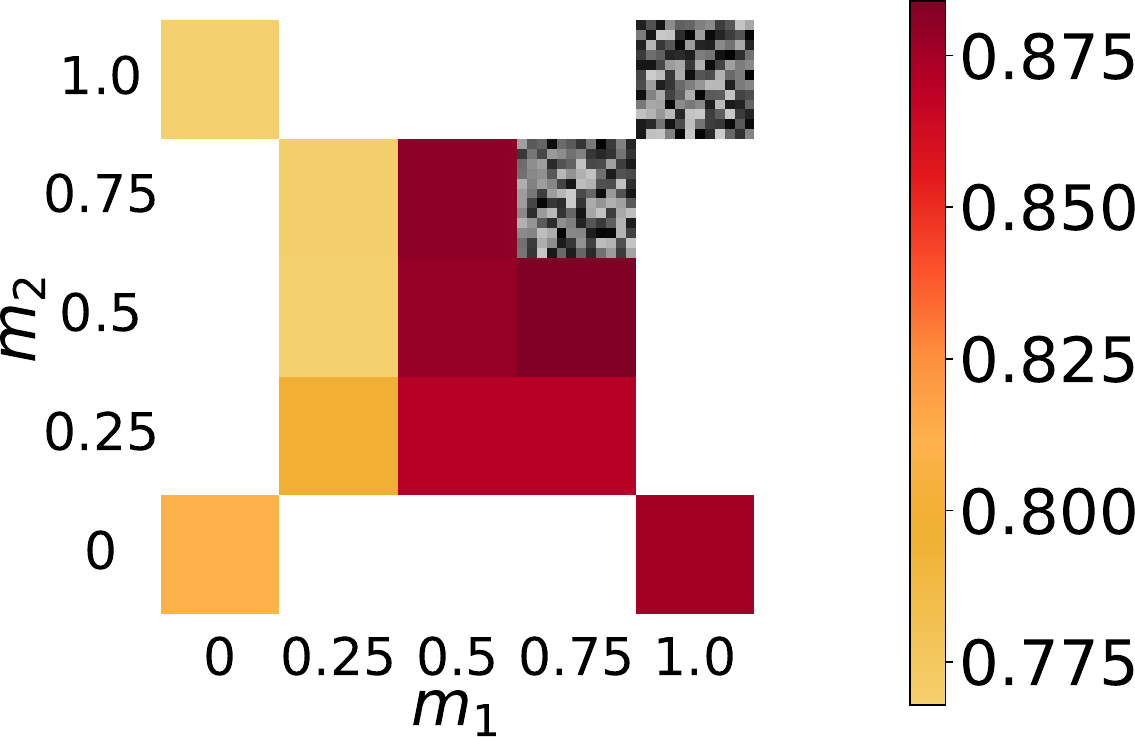} 
        \subcaption{Reconstruction} 
        \label{subfig:1}       
    \end{subfigure}
    \hfill 
    \begin{subfigure}{0.22\textwidth}
        \centering
        \includegraphics[width=\linewidth]{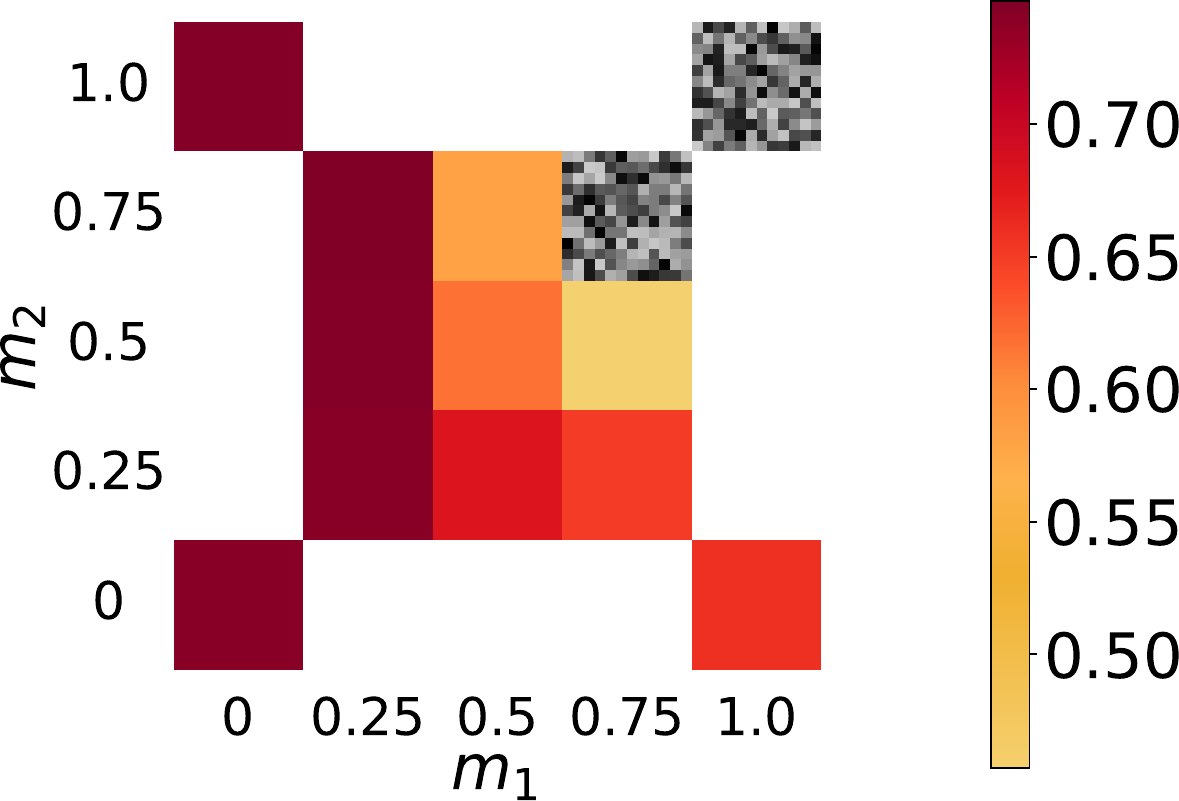}
        \subcaption{Understanding}
        \label{subfig:2}
    \end{subfigure}
    \begin{subfigure}{0.22\textwidth}
        \centering
        \includegraphics[
        width=\linewidth]{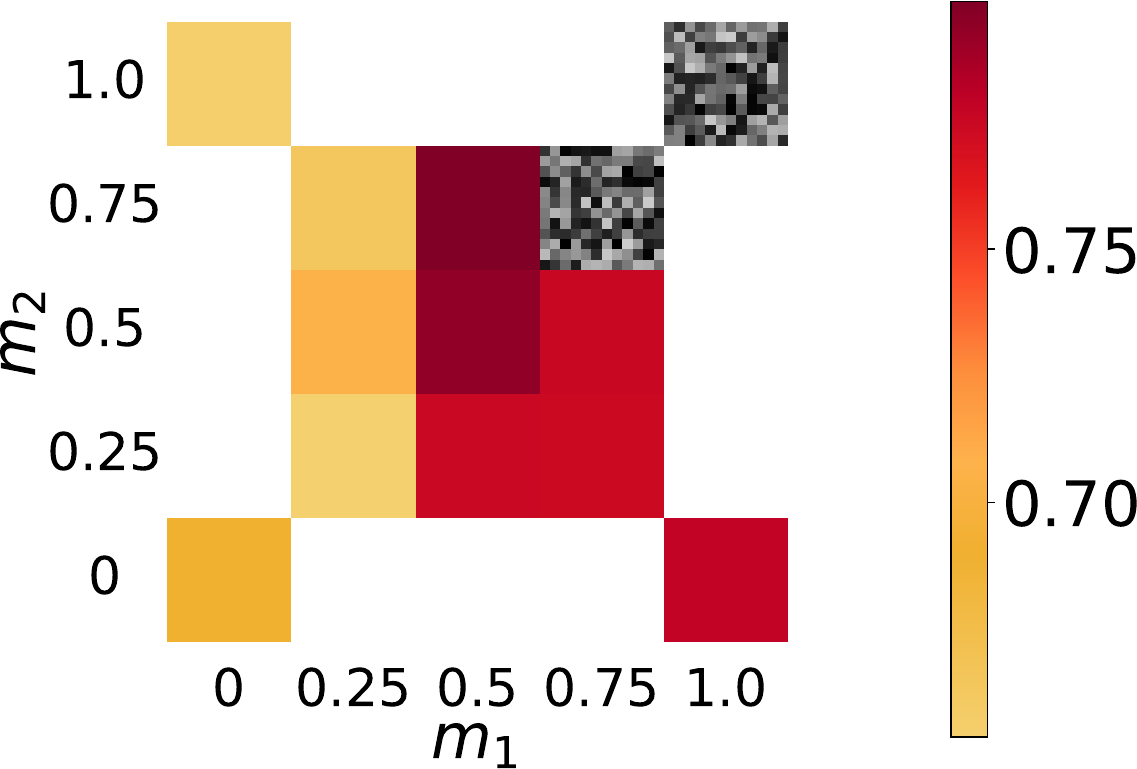}
        \subcaption{Generation}
        \label{subfig:2}
    \end{subfigure}
    \hfill
    \begin{subfigure}{0.22\textwidth}
        \centering
        \includegraphics[width=\linewidth]{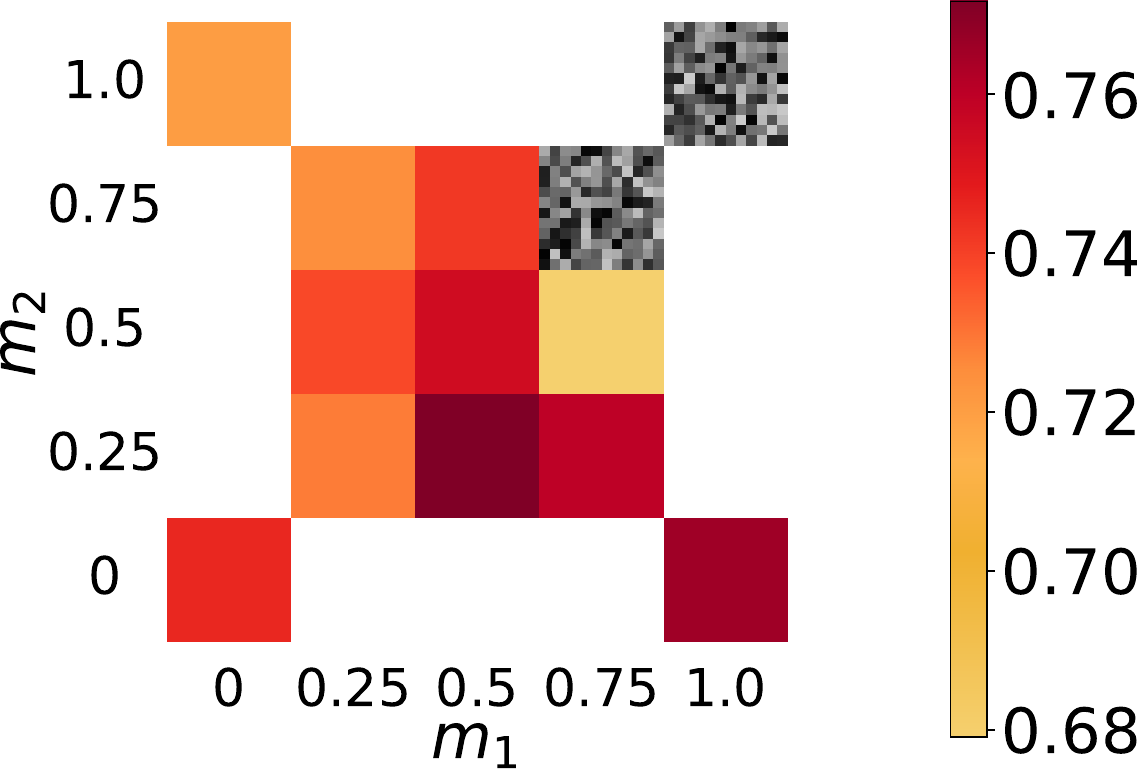}
        \subcaption{Overall}
        \label{subfig:2}
    \end{subfigure}
    \begin{subfigure}{0.22\textwidth}
        \centering 
        \includegraphics[width=\linewidth]{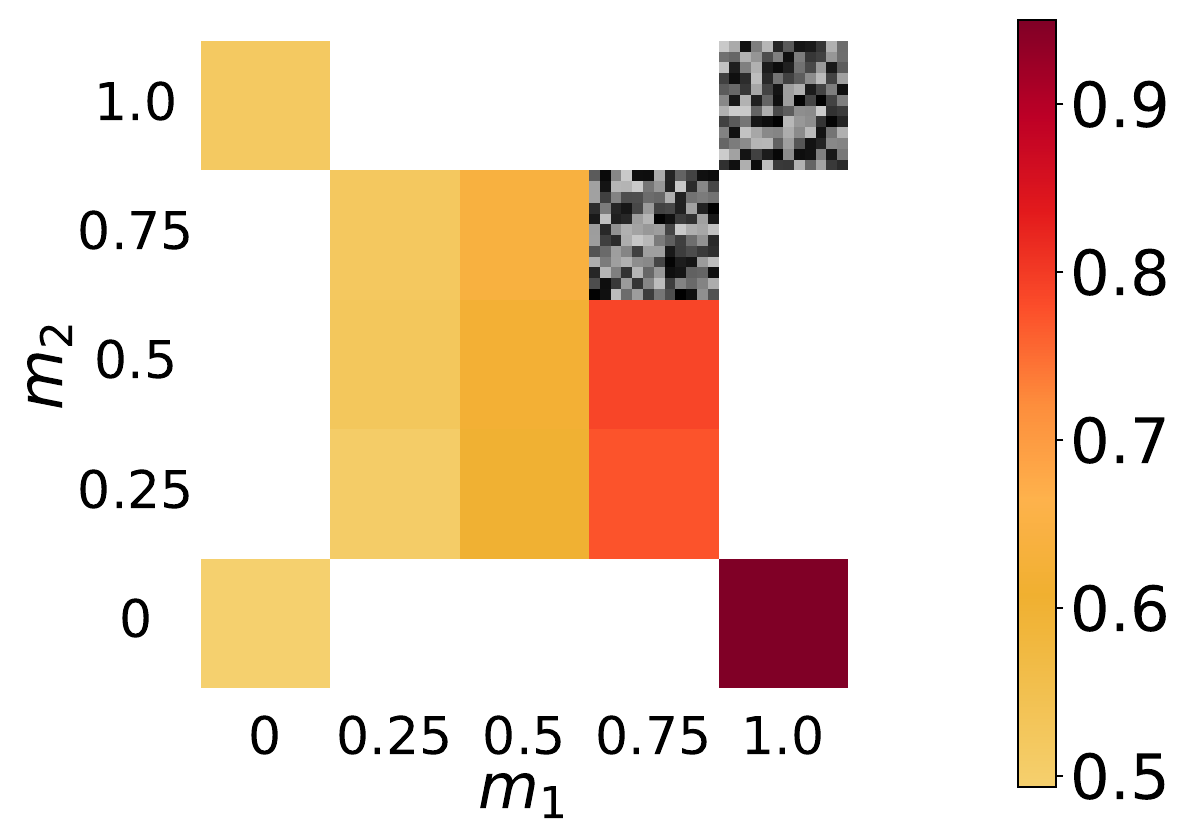} 
        \subcaption{mcos distance (Eq. \ref{mcos})}
        \label{subfig:1}      
    \end{subfigure}
    \hfill 
    \begin{subfigure}{0.22\textwidth}
        \centering
        \includegraphics[width=\linewidth]{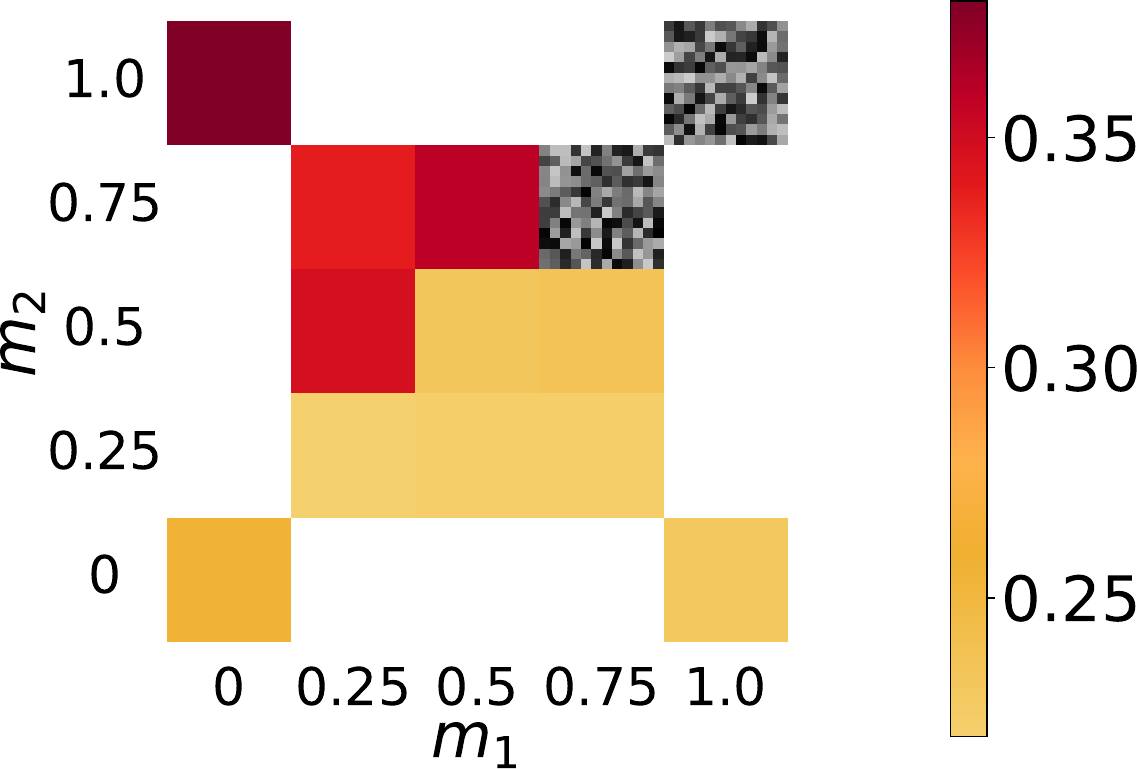}
        \subcaption{mdss distance (Eq. \ref{mdss})}
        \label{subfig:2}
    \end{subfigure}
    \caption{Reconstruction, understanding, generation, and overall performance as well as marginal loss distances for different margin parameter combinations. Mosaic patterns indicate that the VAE training has diverged.} 
    \label{fig:grid_search}        
\end{figure}

Fig. \ref{fig:grid_search} shows the reconstruction, understanding, generation and overall scores of adaptive-weighted JMAS-VAE under various margin combinations $(m_1, m_2)$, as well as the corresponding distances between VAE latents and SSL features calculated by modified Eqs. \ref{mcos} and \ref{mdss} (with $m_1 = m_2 =0$).
Comparing subplots (a)--(d), we can see that \textbf{smaller margins generally improve understanding but impair reconstruction and generation}.
Overall, the heatmap shows no strong symmetry between margins: $(m_1 = 1, m_2 = 0)$ performs much better than $(m_1 = 0,m_2 = 1)$ in reconstruction and generation, revealing the distinct contributions of the two marginal losses.

Subplots (e) and (f) illustrate the two distances between VAE and WavLM features under different margin combinations, computed on the LibriSpeech-test-clean dataset.
The representation distances are linked to distillation losses defined in Eqs. \ref{mcos} and \ref{mdss}, where a smaller distance reflects the optimization direction of distillation
It can be readily observed that a larger margin setting leads to a correspondingly larger loss distance.
Furthermore, Tab. \ref{tab:correlation_distance_score} shows the Pearson Correlation Coefficients (PCC) between representation distances and their correlations with reconstruction, understanding, generation and overall scores for the 11 JMAS-VAE experiments in Fig.~\ref{fig:grid_search}.
It reveals that smaller $\mathcal{L}_{\text{mcos}}$ distances (which are essentially T-axis distances)  strengthen understanding and TTS text accuracy but weaken reconstruction and TTS SIM, thus negatively affecting the overall score.
For the $\mathcal{L}_{\text{mdss}}$ distance, however, the opposite is true: its optimization direction is consistent with the improvement of reconstruction and TTS SIM.

The above observation indicates that the frame-wise alignment scheme on the T-axis focuses more on semantic information, while reducing $\mathcal{L}_{\text{mdss}}$ allows for better preservation of acoustic information.
Carefully balancing the two loss terms with distinct roles eventually leads to a high-quality compact representation for unified reconstruction, understanding, and generation.

\begin{table}[h]
  \centering 
  \fontsize{8}{9.5}\selectfont
  \setlength{\tabcolsep}{2pt} 
  \renewcommand{\arraystretch}{1.2} 
  \caption{Pearson Correlation Coefficients (PCCs) between distance metrics calculated on LibriSpeech-test-clean and task performance scores.} 
  
  \begin{tabular}{c|c|c|ccc|c} 
    \hline
    \multirow{2}{*}{\textbf{Distance}} & \multirow{2}{*}{\textbf{Recon.}} & \multirow{2}{*}{\textbf{Under.}} & \multicolumn{3}{c|}{\textbf{Gene.}} & \multirow{2}{*}{\textbf{Overall}} \\
    \cline{4-6} 
    & & &\textbf{1-WER} & \textbf{SIM} & \textbf{Overall} & \\
    \hline 
    $\mathcal{L}_{\text{mcos}}$    &     0.723   &   -0.615    &   -0.552    &  0.701       &    0.694 & 0.133      \\
    $\mathcal{L}_{\text{mdss}}$    &     -0.585     &   0.284    &  0.391 & -0.384     &  -0.378       & -0.267         \\
    \hline 
  \end{tabular}
  \label{tab:correlation_distance_score}
  \vspace{-3pt}
\end{table}

\vspace{-6pt}
\section{Conclusion}
\vspace{-3pt}
In this work, we explore the design space of distillation loss functions for speech VAEs aligned with speech foundation models.
Through extensive experiments, we demonstrate that speech VAEs equipped with joint-marginal loss and adaptive weighting can achieve balanced and superior overall performance across reconstruction, understanding, and generation tasks.
In future work, we aim to further explore the design space of speech VAEs in other aspects (e.g., channel dimension, frame rate, etc.) and their applications in Speech LLMs.

\section{Acknowledgments}
We thank Zhikang Niu for valuable discussions during this work.
\section{Generative AI Use Disclosure}
Generative AI is only used for text polishing.
\bibliographystyle{IEEEtran}
\bibliography{mybib}

\end{document}